\documentstyle[aps,prb,epsf]{revtex}

\begin{document}
\draft

\title{  Full-Potential LAPW calculation of electron momentum density
              and related properties of Li}

\author{ Tunna Baruah, Rajendra R. Zope and Anjali Kshirsagar \footnote{
 Author to whom all correspondence should be directed.}}

\address{ Department of Physics, University of Pune, Pune-411007, India}

\date{\today}

\maketitle

\begin{abstract}

        Electron momentum density and Compton profiles in Lithium 
along $\langle 100 \rangle$, $\langle 110 \rangle$, and $\langle 111 
\rangle$ directions are calculated using Full-Potential Linear Augmented 
Plane Wave basis within generalized gradient approximation. The profiles 
have been corrected for correlations with Lam-Platzman formulation
using self-consistent charge density.
The first and second derivatives of Compton profiles are studied to 
investigate  the Fermi surface breaks.  Decent agreement is observed 
between recent experimental and our calculated values. Our values for
the derivatives are
found to be in better agreement with experiments than earlier theoretical
results. Two-photon momentum density and one- and two-dimensional angular 
correlation of positron annihilation radiation are also calculated
within the same formalism and including the electron-positron enhancement
 factor. 

\end{abstract}



\def\therefore{_. {^.} _. \quad \quad}
\def\because{{^.}{_.}{^.} \quad \quad}

\def\lrarro{\Leftrightarrow}

\def\vr{\vec r}

\def\vk{\vec k}

\def\vK{\vec K}

\def\vx{\vec x}

\def\vu{\vec u}

\def\vrp{{\vec r}\,'}

\def\vrpp{{\vec r}\,''}

\def\vk{\vec k}

\def\cg{ C_{\vG}}

\def\vtau{\vec \tau}

\def\vp{\vec p}

\def\vpp{{\vec p}\,'}

\def\allvr{\{\vr_j\}}

\def\tpmd{\rho^{2\gamma}}

\def\etal{{\it et al. }}

\def\vnabla{{\vec \nabla}}
\def\vnablau{{\vec \nabla}_{\mu}}

\def\stop{\nonumber   \\}

\def\vG{\vec G}
\def\vg{\vec G}
\def\vGp{{\vec G}\,'}
\def\vRpp{{\vec R}\,''}

\def\vkg{\vk +\vG}

\def\vkt{{\vec k}t}

\def\dE{\Delta E}
\def\dt{\Delta t}
\def\cu{C_{\mu}}
\def\jqvk{J(q,\vk)}
\def\jqvkia{J_{IA}^{(0)}(q,\vk)}
\def\jqvko{J^{(1)}(q,\vk)}
\def\jqvkt{J^{(2)}(q,\vk)}

\def\jone{J^{(1)}}
\def\jtwo{J^{(2)}}
\def\jzero{J^{(0)_{IA}}}
\def\jcumul{J^{(0)_{total}}}

\def\ketphi{\mid \Phi_i \rangle}
\def\braphi{\langle \Phi_i \mid}

%
\def\emiktq{e^{-iktq}}
\def\eitxcu{e^{it(X+\cu)}}
\def\eitxcup{e^{it'(X+\cu)}}
\def\eitxcupp{e^{it''(X+\cu)}}
\def\eitcu{e^{it\cu}}
\def\eitcup{e^{it'\cu}}
\def\eitcupp{e^{it''\cu}}
\def\emitcu{e^{-it\cu}}
\def\emitcup{e^{-it'\cu}}
\def\emitcupp{e^{-it''\cu}}

\def\emiqr{ e^{-iqR}}
\def\emiqR{ e^{-iqR}}
\def\ergmu{ e^{ \vR \cdot \vnabla_\mu}}
\def\ergmup{ e^{ \vR' \cdot \vnabla_\mu}}
\def\ergmupp{ e^{ \vR'' \cdot \vnabla_\mu}}
\def\emrgmu{ e^{ - \vR \cdot \vnabla_\mu}}
\def\emrgmup{ e^{ - \vR' \cdot \vnabla_\mu}}
\def\emrgmupp{ e^{ - \vR'' \cdot \vnabla_\mu}}


\def\rmu{\vr_{\mu}}
\def\vrmu{\vr_{\mu}}
\def\frmu #1{ #1(\ldots,\vrmu,\ldots)}
\def\fnrmu #1 #2{ #1(\ldots,\vrmu+#2,\ldots)}
\def\frmuall #1{ #1(\vr_1,\ldots,\vrmu,\ldots,\vr_N)}
\def\fn #1 #2{ #1(\vr_1,\ldots,\vrmu+#2,\ldots,\vr_N)}

\def\dalln{ d^{3N}r}

\def\kbpi{\frac{k}{2\pi}}
\def\onebpi{\frac{1}{2\pi}}
\def\summu{\sum_{\mu=1}^{N}}
\def\Summu{\sum_{\mu=1}^{N}}
\def\intii{\int_{-\infty}^{\infty}}
\def\intzi{\int_{0}^{\infty}}

\def\fmutvk{F_{\mu}(t,\vk)}
\def\fmuvkt{F_{\mu}(t,\vk)}
\def\fmuvkto{F_{\mu}^{(1)}(t,\vk)}
\def\fmuvktt{F_{\mu}^{(2)}(t,\vk)}
%

\def\vks{v}
\def\psiks{\psi}
\def\psiksc{\psi^*}
\def\dmone{\Gamma^{(1)}(\vr\mid\vr+\vR)}

\def\rhat{\hat r}

\def\phivR{\Phi_{\vR}}


\def\wxc{ W_{xc}}
\def\ex{ {\vec {\cal{ E}}}_{x}}
\def\exr{ {\vec {\cal{ E}}}_{x}(\vr)}
\def\exc{ \vec {\cal {E}}_{xc}}
\def\ee{ \vec {\cal {E}}} 


\newcommand {\grrp}{ {\Gamma}^{(1)} (\vec{r},\vec{r'})}
\newcommand {\gssp}{ {\Gamma}^{(1)} (\vec{s},\vec{s'})}
\newcommand {\gsspnv}{ {\Gamma}^{(1)} (s,s)}
\newcommand {\grr}{ {\Gamma}^{(1)} (\vec{r},\vec{r})}
\newcommand {\grrr}{{\Gamma}^{(1)} (r,r')}
\newcommand {\gppv}{ {\Gamma}^{(1)}_{mom} (\vec{p},\vec{p'})}
\newcommand {\gppp}{{\Gamma}^{(1)}_{mom} (p,p')}
\newcommand {\gpvp}{{\Gamma}^{(1)}_{mom} (\vec{p},\vec{p'})}
\newcommand {\gp}{{\gamma}(\vec{p})}


\newcommand {\dmrrp}{{\gamma}(r,r')}
\newcommand {\Dmone}{{\Gamma}}
\def\ijN{1{\le}i<j{\le}N}
\def\modrrp{ \mid\vr-\vrp\mid}
\def\half{ \frac{1}{2}}
\def\spinj{ \sum_{j \atop {{\rm spin}\, i= {\rm spin}\, j}}}
\def\spinij{ \sum_{ij \atop {{\rm spin}\, i= {\rm spin}\, j}}}
\def\Minden{ {{\rm Min} \atop {\scriptstyle{\Psi_n\Rightarrow n(\vr)}}} }
\def\Mindeno{ {{\rm Min} \atop {\scriptstyle{\Psi_{n_0}\Rightarrow {n_0}(\vr)}}} }
\def\density{ n(\vr)}
\def\nup{ n_{\uparrow}(\vr)}
\def\ndn{ n_{\downarrow}(\vr)}
\def\Xa{ X_{\alpha}}

\def\hT{\hat T}
\def\hU{\hat U}
\def\hH{\hat H}
\def\hHp{\hat H'}
\def\hVp{\hat V'}
\def\hvp{\hat v'}
\def\hV{\hat V}
\def\hv{\hat v}

\def\IR{ I\!\!R}
\def\II{ 1\!\!I}
\def\hp{ \hat p}

\section{ Introduction}

         Compton scattering and positron annihilation techniques
are well established tools for studying the momentum distribution 
of electrons in solids \cite{BW,Singru,West,berko}.  In the Compton 
scattering technique, the intensity distribution of energy broadened
Compton scattered radiation (called the Compton profile) is studied
while in positron annihilation technique, measurement of angular
correlation between two photons emitted during the annihilation of a
thermalized positron with electrons in the solid, is performed.
The Compton profile (CP) and angular correlation of positron annihilation
radiation (ACPAR) curves contain the finger prints of  the Fermi surface 
(FS) breaks in the momentum distribution in the first and higher 
Brillouin zones.  It is well known that although the magnitude of the 
discontinuity in momentum distribution itself changes due to 
electron-electron, electron-ion and electron-positron correlations, 
the position of the discontinuity remains unchanged 
\cite{majumder,migdale,daniel}.
Thus, these techniques together are useful to extract information 
about FS geometry and to identify electron correlation effects.
 Positron annihilation techniques are more sensitive to 
the outer, weakly bound conduction electrons and are also capable of 
performing measurements in both one-dimensional and two-dimensional
geometries, with much superior momentum resolution. Theoretically,
 however, it is more straight forward to calculate CP's as ACPAR 
studies necessitate to account for electron-positron many-body 
correlation effects. These effects are incorporated in the form of 
momentum, energy or density  dependent enhancement factors 
\cite{kahana,mijnarends,puska,jarlborg,daniuk}.

 Early measurements of Compton profile  were 
suffering from the limited momentum resolution in the 
Compton scattering experiment $(\sim 0.4 a.u.)$.  The advent of high 
intensity, high energy and well-polarized synchrotron sources  and 
 the spectrometers  with high resolution  $( \sim 0.12 a.u.)$ have 
resulted in revival  of interest in this area \cite{sakurai,synchro,schulke}. 
On the theoretical side, high performance computing facilities have
made it possible to perform calculations on a fine $\vp$-mesh
with a better convergence criteria for the total energy and charge 
density self-consistency.

In the present communication, we report full potential
linearized augmented plane wave (FP-LAPW) calculations of the 
electron momentum distribution in $Li$.  In lithium, because of 
its smallness, the electron-ion interactions are strong and the 
electrons do not behave like a text-book example of homogeneous 
electron gas as in sodium. The Fermi surface of $Li$ shows a small 
but definite departure from free electron sphere. The electron 
momentum density is highly anisotropic, the high momentum components 
(HMCs) are small but important.  The momentum distribution of  $Li$ 
has been investigated in the past, both theoretically and experimentally,
by several workers \cite{Lafon,Lundqvist,Eisenberg,Yamashita,Calaway} with the 
available state-of-the-art procedures. Recently, Sakurai \etal 
\cite{sakurai}  have performed high resolution Compton scattering 
experiments for $Li$ to measure Compton profiles and Fermi radii.  The 
measured CP's are compared with the theoretical ones calculated using 
KKR method. Subsequently, Sch\"ulke and coworkers  \cite{schulke} measured 
$11-$directional Compton profiles and employed them to reconstruct the three 
dimensional EMD in $Li$ metal. The comparison of the experimental 
CP's with their theoretical counterparts  have always shown some 
discrepancies.  This is partly due to the various approximations 
involved in computing the profile and partly due to the experimental 
errors. The methods like FP-LAPW can provide accurate results for the 
one-electron wavefunction and hence should be employed for computation of 
theoretical CP's.  Such accurate calculations of the CP's of $Li$ 
have been reported by Kubo \cite{kubo1,kubo2} employing the GW 
approximation using  FP-LAPW wavefunctions within local density 
approximation (LDA) as the zeroth approximation. 
 The calculated CP's are in good agreement  with the experimental values 
but the derivatives of the CP's match the experimental ones poorly 
\cite{kubo1}. The derivatives of CP's are important as they provide 
information about the Fermi surface. 

In the present paper, we use full-potential linearized augmented plane 
wave method for the computation of CP which have been corrected to include
the correlation effects not accounted for within LDA, alongwith one and 
two dimensional
 angular correlation of positron annihilation radiation (1D- 
and 2D-ACPAR) which include corrections due to  electron-positron 
correlation effects. Such studies using the same formalism for 
calculation of CP and ACPAR can provide complementary information 
about the Fermi surface and the electron momentum distribution.
Our aim in this communication is to extract the FS geometry from the 
two techniques together and to identify the part of electron 
correlations left out in the theory to describe the EMD.
The plan of the paper is as follows: in section II, we present  
the momentum space formulation of the LAPW wavefunction  and 
computational details, while section III deals with the results.

\section{ Methodology and computational details}

        The one electron wavefunction for an electron in the state
labeled by wave vector $\vk$ and band index $j$ is expanded in the 
Linearized Augmented Plane Wave (LAPW) basis $\phi_{\vkg} (\vr)$ as
\begin{eqnarray}
         \psi_{\vk}^j (\vr) = \sum_{\vg} c_{\vkg}^j  \phi_{\vkg} (\vr) ,
                      \label{eq:rwfn}
\end{eqnarray}
 where   $c_{\vkg}^j$ are the expansion coefficients and $\vG$ denotes
reciprocal lattice vector.
 The LAPW function is a plane wave outside  the MT 
sphere 
while inside it is a linear combination 
of  $u_l(r)$, the solution of the radial Schr\"odinger equation and its
 energy derivative $ \dot{u}_l(r)$. This allows
greater flexibility  inside the spheres (than the APW method) and 
hence permits  computation of an accurate solution\cite{Dsingh}.  

The momentum space  LAPW wavefunction is obtained by a 
Dirac-Fourier transformation  of eq. (\ref{eq:rwfn}) and
the electron momentum density is computed from the momentum space 
wavefunction of occupied states.
The Compton profile is then obtained by performing a double 
integral. 

 The theoretical Compton profile thus obtained is overestimated at low 
momenta and is underestimated at higher momentum values   than its 
experimental counterpart. This discrepancy  is often attributed to the 
correlation effects that are ignored in the independent particle model\cite{LP}.   
Lam and Platzman \cite{LP} have shown that these effects can be 
incorporated into the EMD by augmenting to the independent electron 
model momentum density 
 a correction term.
 The recent formulation of Cardwell and Cooper \cite{cardwell} based 
on the Lam-Platzman work which takes care of the non-unity occupation below
the Fermi momentum $k_f$ and non-zero occupation beyond $k_f$
has been employed in the present work.

For a single positron in a defect free crystal, the positron density
will be distributed over the entire crystal. Therefore,
the  positron  wavefunction, $\psi_+(\vr)$, is considered to be delocalized
 and is represented by a plane wave basis. Further, as the positron 
essentially thermalizes before annihilation, it is assumed to be in 
the state $\vk_+=0$. The enhancement in the TPMD due to the
electron-positron short-range correlation is calculated according to 
the prescription given by Puska \etal \cite{puska} which is a 
function of electron and positron densities.  Thus, TPMD is evaluated 
using the following prescription :

\begin{equation}
\rho^{2 \gamma} (\vp) 
= \sum_{\vk,j}^{occ}  \mid F^j_{\vk} (\vp) \mid^2  
 = \sum_{\vk,j}^{occ} \mid \int e^{-i {\vp.\vr}} \psi_+ (\vr)
\psi_{\vk}^j (\vr)  \sqrt{ g (n^e(\vr), n^+(\vr))} ~~d^3r ~~ \mid^2 . 
\end{equation}
where $n^e(\vr)$ and $n^+(\vr)$ are the electron and positron densities 
respectively
and the enhancement factor $g(n^e(\vr),n^+(\vr))$  
in the limit $n^+(\vr) \longrightarrow 0$ is
\begin{eqnarray}
g_0(r_s)=1 + 1.23r_s + 0.98890 r_s^{3/2} - 1.4820 r_s^2 + 0.3956 r_s^{5/2}
+ r_s^3/6 
\end{eqnarray}
where $ r_s = \biggl [\frac{3}{4 \pi n^e(\vr)} \biggr ]^{1/3}$.

   A self-consistent band-structure calculation was performed using the 
LAPW method as implemented in WIEN97 package \cite{blaha}.
The calculation employs full potential which implies that the
nonspherical part of the potential inside the muffin-tin sphere and its
deviation from the constant potential in the interstitial region are
taken into consideration. The simplified generalized gradient 
approximation (GGA) due to Perdew \etal \cite{PBE} was used for the
exchange-correlation part of the Kohn-Sham potential.
 In the band structure calculation the lattice constant of $Li$ in
 bcc structure is taken to be $ 6.61375 a.u.$ \cite{crc}. 
The self-consistency cycles were carried out to an energy tolerance of
$10^{-6}$ Ryd and charge convergence of $10^{-5}$ electrons. The various
band-structure parameters agree very well with earlier accurate
calculations \cite{Calaway,moruzzi}.

For the calculation  of Compton profile, in  1/48th of the 
Brillouin zone (BZ) we used 40425 ${\vk}$-points to evaluate the 
momentum space wavefunction which when translated by the
reciprocal space vectors yield $40425 \times 531$ ${\vp}$-points. 
Linear tetrahedron method \cite{lehman}  was used for the calculation 
of Compton profiles over a momentum mesh of $ 0.001 a.u.$.  The correlation 
correction was carried out as described by Cardwell {\it et al.} 
\cite {cardwell}. In carrying out these calculations we
used the self-consistent density inside the MT sphere whereas in the
interstitial region we assumed the density to be flat with little
structure. No $r_s$ cut-off value was used as suggested by Cardwell
\etal \cite{cardwell}.

 For the calculation of two photon momentum density, 
the wavefunction for the positron was obtained by solving the secular 
determinant once using the self-consistent Coulombic potential from 
earlier calculation for electronic band structure but with an opposite 
sign.  We have used 819 $\vk$-points to evaluate $F_{\vk}^j(\vp)$.

\vspace{0.2in}
\section{ Results and discussion }

\vspace{0.1in}

The electron momentum density for lithium metal along $<100>$, $<110>$ 
and $<111>$ directions in the momentum space is plotted in Fig. 1. Each 
panel shows EMD along one direction, calculated
within LDA and GGA employing von Barth-Hedin (VBH) \cite{vbh} 
 and Perdew-Burke-Ernzerhof (PBE) \cite{PBE} exchange-correlation 
potentials respectively alongwith
 GGA-EMD corrected using Lam-Platzman (LP) 
\cite{LP} correction term in that direction employing
the model momentum density proposed by Cardwell \etal \cite{cardwell}. 
The results are almost identical for VBH and PBE exchange-correlation
potentials indicating that the non-local corrections as described
within GGA do not seem to affect the momentum density in $Li$
although the electronic structure is slightly different in LDA and GGA. 
This gives rise to almost identical Compton profiles (CPs) for LDA and
GGA formulations as predicted by Lam and Platzman \cite{lam2}.
These observations for $Li$ support the fact that although the ionic
potential is strong in $Li$, the conduction electron density behaves
more like a homogeneous electron gas. Similar calculations for
transition metals do show significant differences in LDA and GGA results
\cite{ni}.

 The strong ionic periodic potential, however, does couple the states
near Brillouin zone (BZ) boundaries and the conduction electron
wavefunctions contain strong high momentum components and the EMD does
show anisotropic behavior as is evident from Fig. 1. Since $Li$ has only
one electron in the conduction band, in the one-electron picture, the
EMD is zero beyond $k_f$ in the first BZ and between its images in higher
zones. Lam-Platzman \cite{LP} correction partly takes care of the correlation
effects on the wavefunctions in one-electron picture. The occupation
number of an interacting homogeneous electron gas is estimated to be
smaller than that of a non-interacting free electron gas by 4\% for $Li$
\cite{daniel}. Our LP corrected EMD displays the effect of states
below $k_f$ being pushed above $k_f$ in $Li$ and these can be compared with
results in Fig. 5 of ref. \cite{Lundqvist} which uses mean occupation 
numbers derived from electron gas data for correlation effects and couples 
them with orthogonalized plane wave band structure method. EMD dies off in 
higher zones along $<100>$ and $<111>$ directions but shows a strong 
umklapp component along $<110>$ direction. These higher momentum components
(HMCs) play an important role in determining the shapes of Compton
profiles and angular correlation curves.

Compton profiles $J_{\hat{k}}(q)$ corrected for the correlations are 
presented in Fig. 2 together with the experimental and KKR results 
\cite{sakurai} with $\hat{k}$ along $<100>$, $<110>$ and $<111>$
 directions. In order to facilitate the comparison with  experimental 
CP's, the theoretical CP's were convoluted with a Gaussian of FWHM 
of 0.12 a.u.  (experimental resolution in ref. \cite{sakurai}). 
Our  Compton profiles, when compared with experimental values, support 
the widely known behavior, namely, overestimation at low momentum and 
underestimation at higher momentum values. Correlation corrections 
lower the CP values within the main Fermi surface but they are still 
higher than the experimental results.  It is to be noted that KKR 
values are consistently higher than our values near $q=0$ and show 
a more pronounced cusp like behavior near $\vk_f$ for all the three 
directions.  The cusp seen in CP reflects the discontinuity in EMD
at $\vk_f$ in the first BZ and their images in higher zones. The results
indicate that the discontinuity is smaller in FP-LAPW calculations 
than in KKR.  The disagreement between present results and KKR results
could be attributed to the fact that present work  is a full potential calculation 
whereas KKR uses muffin-tin shape approximation. Secondly, the LP correction
has been calculated using the prescription of Cardwell {\it et al.}
\cite{cardwell}
employing self-consistent charge density
 in the present work whereas KKR uses interpolated 
results from homogeneous electron gas data.  Our results agree with those 
of  KKR beyond $q=0.8~ a.u.$ while the experimental values are consistently 
higher than the theory in this region.  This reflects that the non-zero 
occupation beyond $\vk_f$ for the inhomogeneous electron gas is only 
partially accounted for by the theory.  

 We have also compared the directional anisotropies in the Compton 
profiles with the experimental work of Sakurai \etal \cite{sakurai} 
and the overall agreement is found to be good. 
 The directional anisotropies are important while comparing theory 
with experiment as the systematic errors in the experimental and 
theoretical results are canceled out. The prominent
structures near $q=0$ are well reproduced at the correct momentum 
values, however, they are overestimated by theory (Fig. 3). 
This is again due to the correlation correction functional
as discussed by Bauer \etal \cite{Bauer}.  In the present calculation, 
we have included the correlation correction which is isotropic and 
therefore does not affect the results of anisotropy. Attempts to 
include anisotropic corrections are in progress and will be published
elsewhere.

The two-dimensional ACPAR surfaces were obtained by integrating $\rho
^{2 \gamma}({\vp})$ along $<100>$, $<110>$ and $<111>$ directions
respectively; while the one dimensional curves were obtained using the 
linear tetrahedron method.  The 1D ACPAR curves along $<100>$, $<110>$ and 
$<111>$ are shown in Fig. 4 and are in qualitative agreement with earlier
published work \cite{Stewart,Meingailis}. The 1D ACPAR curves were 
convoluted with a Gaussian with FWHM of 0.022 a.u.. The 2D 
ACPAR data convoluted with 0.5x0.23 mrad$^2$ FWHM \cite{prasad} is 
presented in Fig. 5 and the overall shape matches with experiment 
\cite{Oberli}.  The higher momentum components in TPMD are seen 
quite clearly in the 2D ACPAR plot. The smaller bump comes from the 
TPMD in the second zone in the $<110>$ direction, whereas the 
contributions from higher zones are hardly visible.  The larger bump 
is a result of projected contributions from $<01{\bar 1}>$ and 
$ <10 {\bar 1}>$  for the same $p$ value. The HMC intensity is seen 
to be a sharp function of momentum values. The momentum density decreases 
with increasing momentum values which is a direct reflection of 
$s$-like wavefunction.

We have extracted relevant FS data from both electron and
electron-positron momentum distributions. Although, in principle, both
 the Compton scattering and the ACPAR probe the electron momentum 
distribution and provide complementary information about FS, 
the latter however provides the best possible measurements for FS breaks in EMD 
since the electron-positron correlations enhance the momentum density 
at $k_f$. The prominent breaks in momentum distribution are seen around 
$0.6~ a.u.$  with slightly different values along different directions as 
shown in Fig. 1 for EMD. Similar structure is seen in TPMD also.  The 
images of the FS breaks, seen in higher zones due to periodicity,
are reflected in the CP and ACPAR data. The structures seen are
identical to those seen earlier and discussed by Sakurai {\it et al.}
\cite{sakurai} for CP and Stewart {\it et al.} \cite{Stewart} for ACPAR
curves.  However, we point out that the LP correction
shifts the FS breaks as seen in the derivatives of CP. This is a 
limitation of the model which employs isotropic momentum density
 to calculate the LP correction.

Figure 6 displays the first derivatives of the directional Compton profiles. 
Although the breaks in the first derivative curve  without convolution 
bring out the structures rather well indicating the distortions of the 
free electron sphere, however,  in Fig. 6, we show the derivatives of
 the convoluted Compton profiles to facilitate comparison with experimental 
results.  The derivatives of 1D ACPAR which bring out the FS breaks sharply 
are presented in Fig. 7. The values of the FS radii along the principal 
symmetry directions $<100>$, $<110>$ and $<111>$  as estimated from the 
positions of peaks in the second derivatives of CP differ from the actual 
Fermi radii as described by Sakurai \etal \cite{sakurai}. The first 
derivatives of directional ACPAR curves give the correct Fermi radii since 
the high momentum components in TPMD are smaller.  Our values of Fermi 
radii alongwith the experimental values \cite{sakurai} are presented 
in Table I. The maximum Fermi surface asphericity, 
$[k_{110}- k_{100}]/k_f^0$ where $k_f^0$ is the free electron radius, 
in our calculation turns out to be 5.6\% against the experimental 
value of 4.6\%, 5\% and 4.7\% respectively obtained from CP 
\cite{sakurai}, 1D-ACPAR \cite{Donaghy} and 2D-ACPAR data \cite{prasad}.

          In this paper, we have presented the Compton profiles 
computed using FP-LAPW method within GGA and have corrected them for 
correlations on the lines of Cardwell and Cooper \cite{cardwell} using
self-consistent charge density.
 The derivatives of the calculated CP's are in good agreement with their
 experimental counterparts. Usually, the discrepancy between theory
and experiment is ascribed to the limitation of local density 
approximation.  However, we have seen that gradient corrections to 
exchange-correlation potential as described in GGA do not affect the 
electron momentum density significantly. In principle, Lam-Platzman 
correction describes the non-local effects on the momentum density 
correctly but the practical implementation is able to account for it 
only partly, namely the isotropic electron correlations are described 
satisfactorily but not the anisotropic ones. Although Bauer \etal 
\cite{Bauer} have rejected the idea of momentum density functional 
theory, we feel a description for the exchange-correlation energy 
functional in momentum space will allow a better quantitative 
representation of EMD theoretically.

 One and two dimensional ACPAR curves are also computed using the 
FP-LAPW method. We have shown that the different shapes of HMCs are 
well reproduced by our calculations. Inclusion of density dependent 
enhancement factor is found to reduce the HMCs at large $p$ values. 
To our knowledge, this is the first theoretical report of 1D- and 
2D-ACPAR for $Li$ incorporating density dependent enhancement effects.

\begin{acknowledgments}

              We gratefully acknowledge the experimental data and KKR 
results provided by Dr. Y. Sakurai and Prof. A. Bansil. We are grateful 
to Prof. K. Schwarz and Dr. P. Blaha for providing the  WIEN97 code. 
We are also thankful to Dr. V. Sundarajan and Prof. R. M. Singru for 
helpful discussions.  TB and RRZ gratefully acknowledge the financial 
support from Council for Scientific and Industrial Research, New Delhi 
in the form of fellowships.

\end{acknowledgments}

\newpage

.

\vspace{0.7in}

\centerline {Table I}

\vspace{0.3in}

\hspace{2.0in}
\begin{tabular}[t]{lcc} \hline
Directions & \multicolumn{2}{c}{ Fermi surface radii}  \\ 
& FP-LAPW & Experimental  \\   \hline 
\\
$k_{100}$ & 0.578  & 0.577 $\pm$ 0.004 \\
\\
$k_{110}$ & 0.611  & 0.604 $\pm$ 0.004 \\
\\
$k_{111}$ & 0.585  & 0.586 $\pm$ 0.004 \\   \hline
\\

\end{tabular}

\newpage

.
\vspace{0.2in}

\begin{figure}[t]
\epsfxsize=6.5in
\epsfysize=8.5in
\centerline{\epsfbox{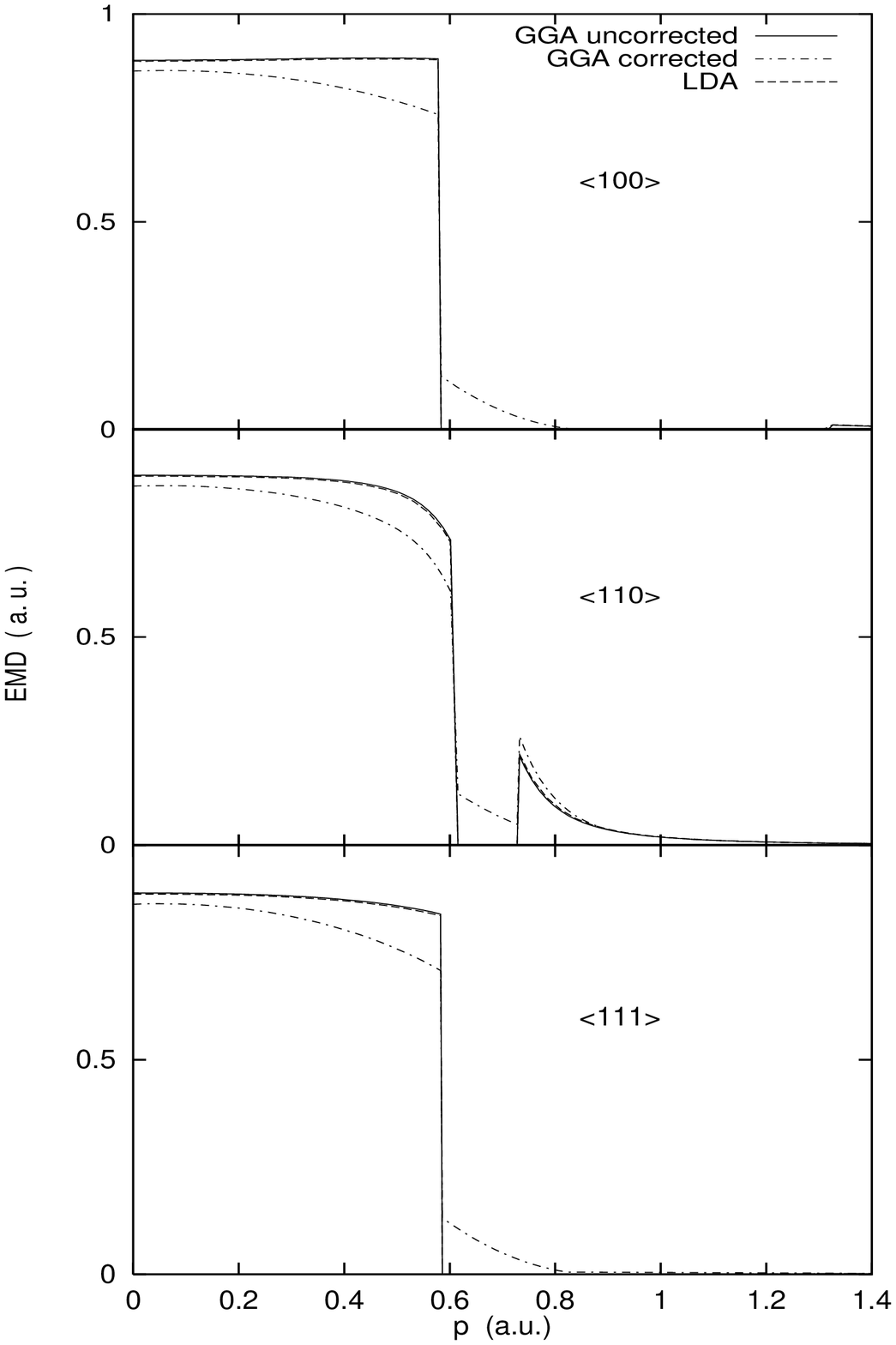}}

\hspace{-0.4in}
\caption { Electron momentum densities along $<100>$, $<110>$ and $<111>$
directions. }
\end{figure}

\newpage

.
\vspace{0.2in}

\begin{figure}[t]
\epsfxsize=6.5in
\epsfysize=8.5in
\centerline{\epsfbox{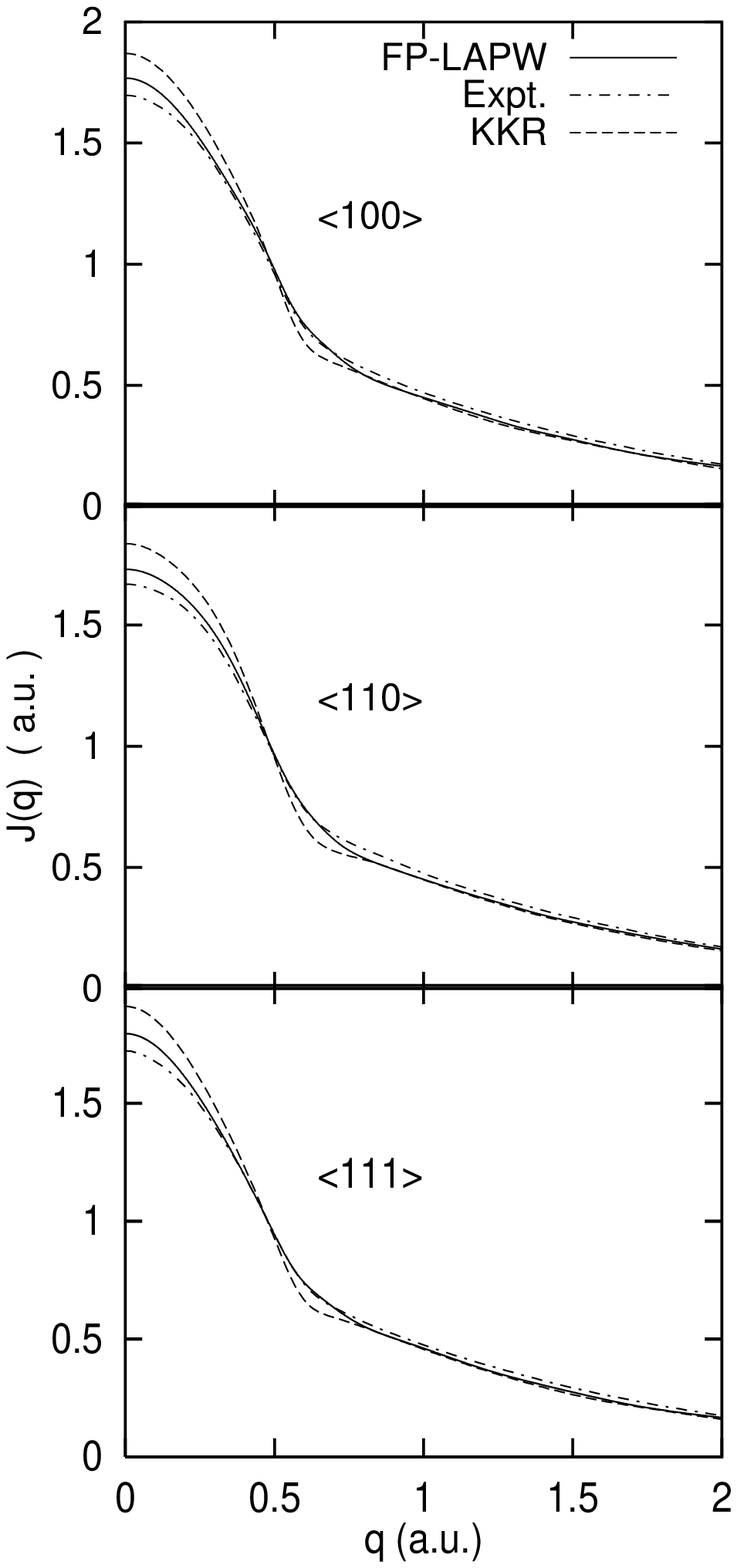}}

\hspace{-0.4in}
\caption { Compton profiles along $<100>$, $<110>$ and $<111>$
directions.}
\end{figure}


\newpage

.
\vspace{0.2in}

\begin{figure}[t]
\epsfxsize=6.5in
\epsfysize=8.5in
\centerline{\epsfbox{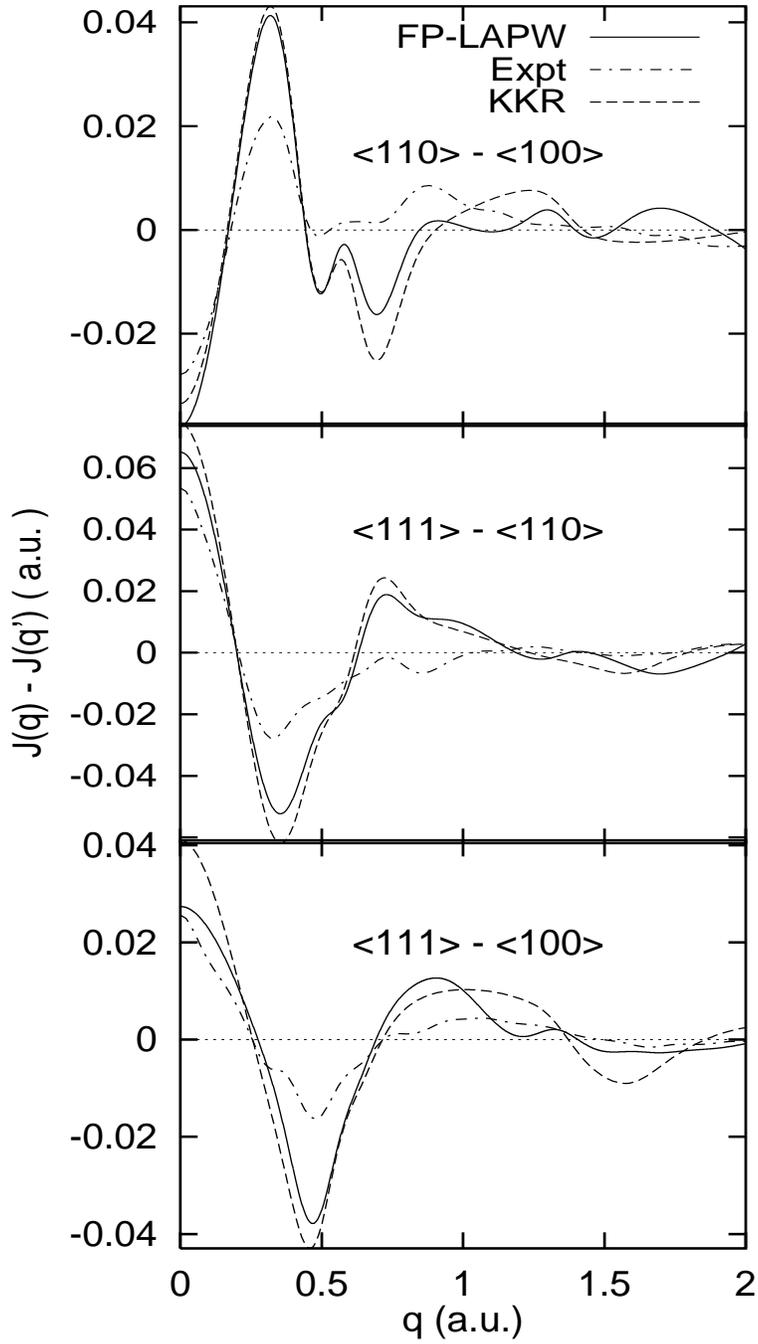}}
\vspace{0.2in}
\caption { Anisotropies of the Compton profiles between various
directions.}
\end{figure}


\newpage
.
\vspace{0.2in}

\begin{figure}
\epsfxsize=6.5in
\epsfysize=8.5in
\centerline{\epsfbox{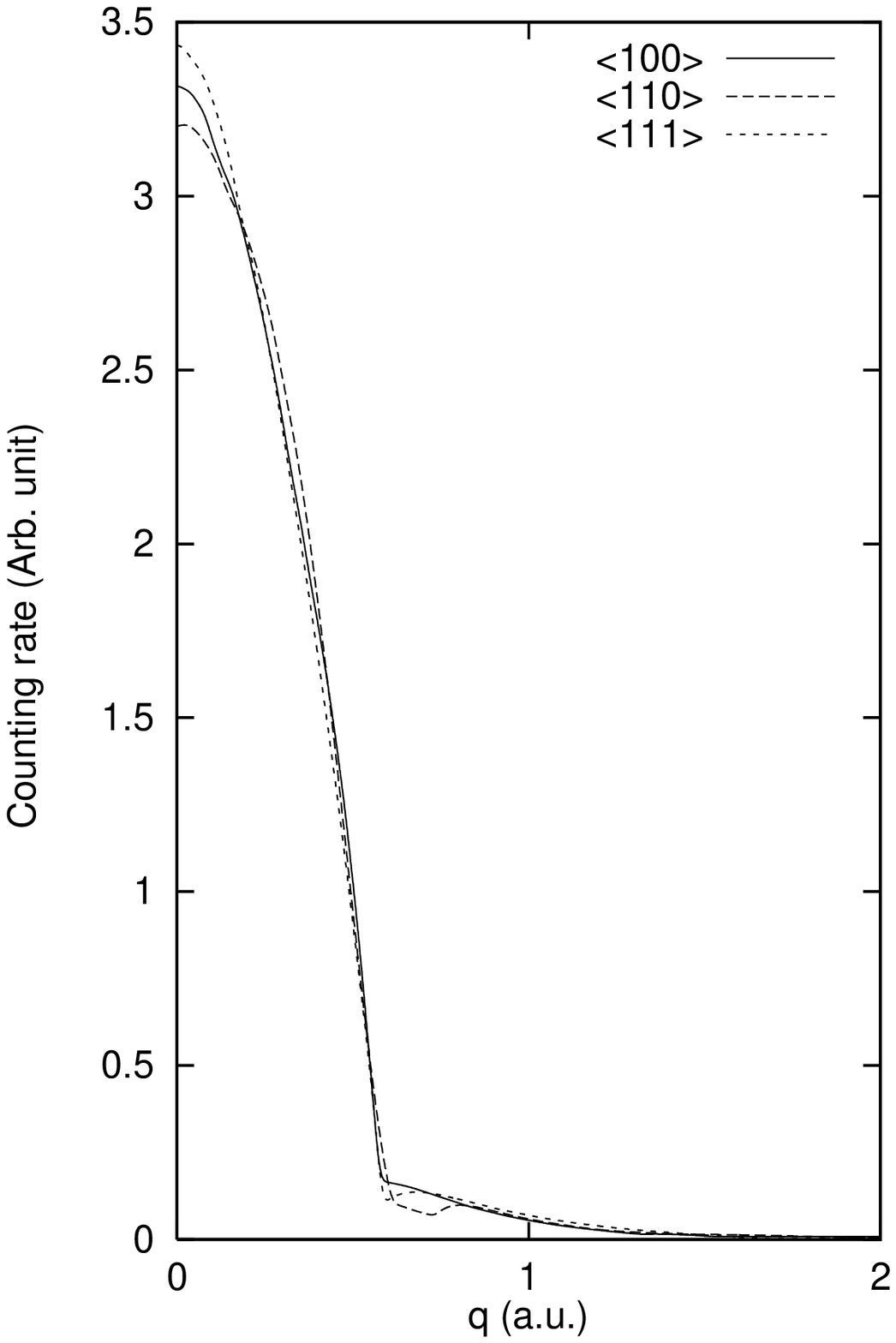}}
\caption{One-dimensional ACPAR curves  along  $<100>$,   $<110>$ and 
  $<111>$ directions }
\label{CPS:fig}

\end{figure}


\newpage

\begin{figure}
\epsfxsize=6.5in
\epsfysize=8.5in
\centerline{\epsfbox{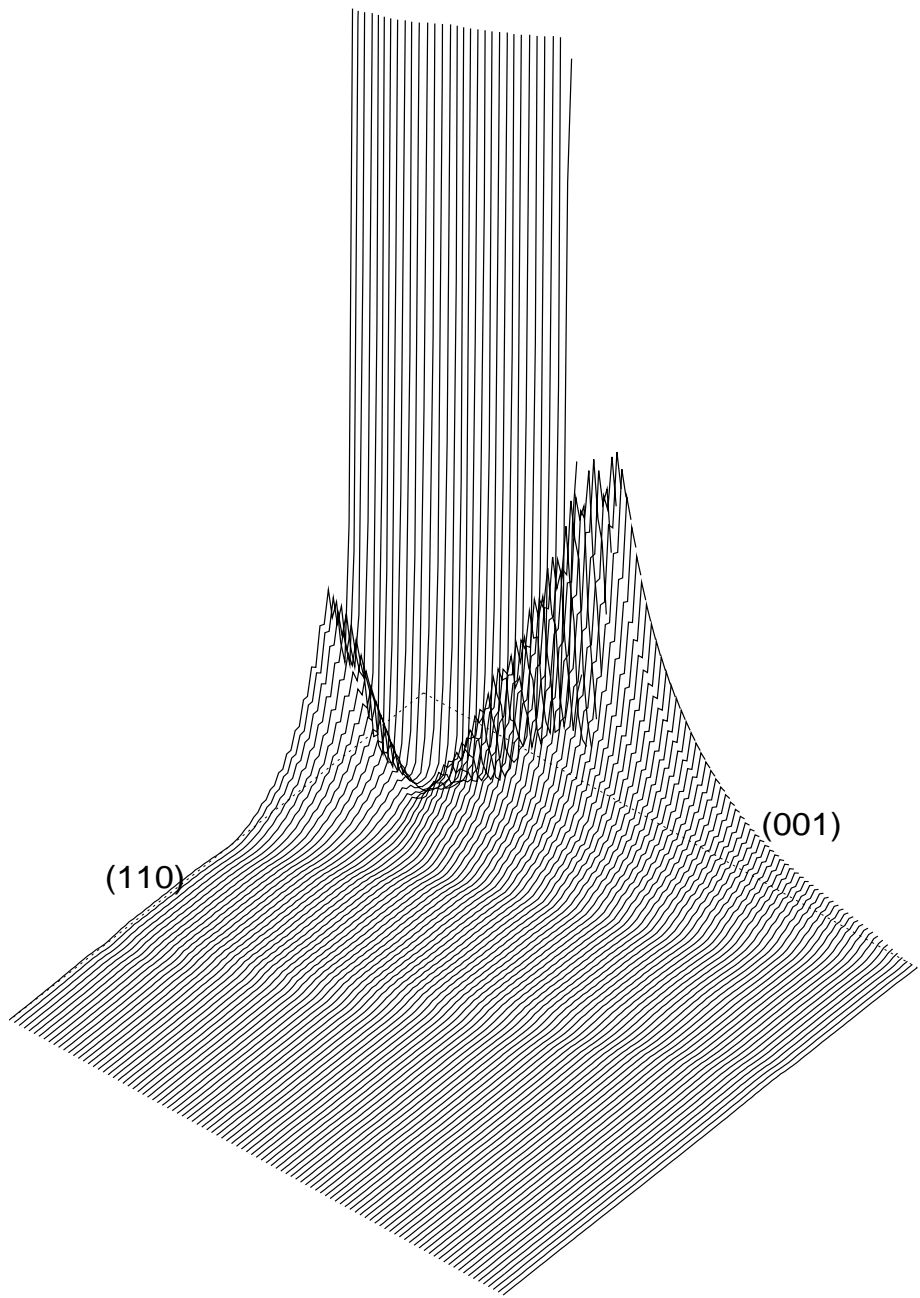}}
\vspace{0.2in}
\caption {Two-dimensional ACPAR plot of $Li$ in the $(1{\bar 1}0)$ 
plane.}
\end{figure}


\newpage

.
\vspace{0.2in}

\begin{figure}
\epsfxsize=6.5in
\epsfysize=8.5in
\centerline{\epsfbox{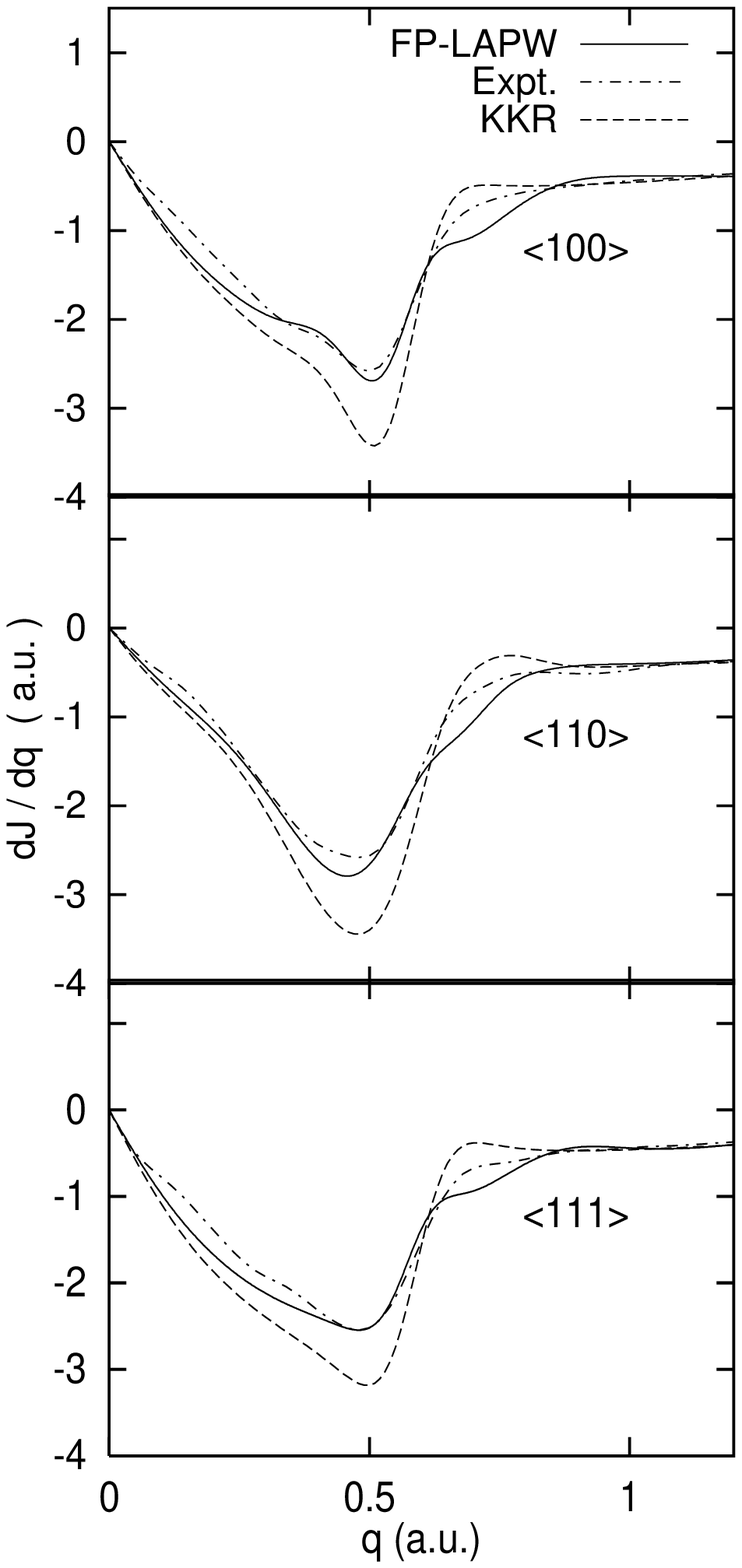}}
\vspace{0.2in}
\caption{ Derivative of the Compton profiles along   $<100>$,  
$<110>$ and  $<111>$ directions.}
\end{figure}


\newpage
.
\vspace{0.2in}

\begin{figure}
\epsfxsize=6.5in
\epsfysize=8.5in
\centerline{\epsfbox{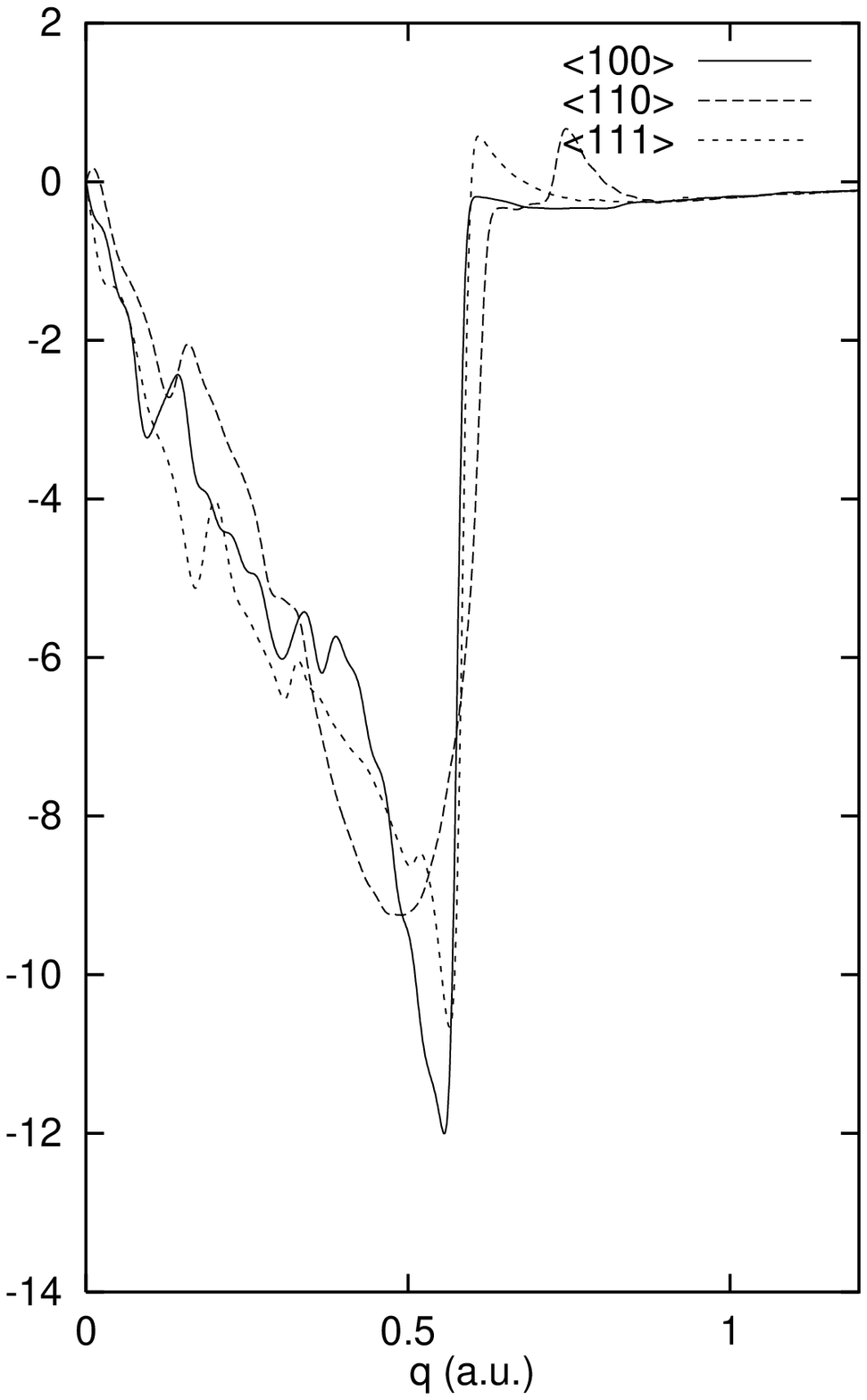}}
\caption{Derivative of one-dimensional ACPAR curves  along  $<100>$,
   $<110>$ and   $<111>$ directions }

\end{figure}


\begin{thebibliography}{99}
\bibitem{ni}
 Our  work on ferromagnetic Ni, which will be published elsewhere,
demonstrates that in Ni, the non local corrections affect the EMD.
This is in agreement with the results in ref. \cite{synchro}.
\end{thebibliography}
\end{document}